# Tunable magnetization in nanoscale LuFeO$_3$: Role of morphology, ortho-hexa phase ratio and local structure

Smita Chaturvedi, [1,2,8*] Priyank Shyam,[3] Mandar M. Shirolkar,[4,5] Swathi Krishna[1,2], Bhavesh Sinha,[6] Wolfgang Caliebe, [7] Aleksandr Kalinko,[7] Gopalan Srinivasan [8] and Satishchandra Ogale[1,2*]

[1]Indian Institute of Science Education and Research, Pune, Dr. Homi Bhabha Road, Pashan, Pune - 411008, India

[2]Centre for Energy Science, Indian Institute of Science Education and Research, Pune - 411008, India

[3]Interdisciplinary Nanoscience Center, Aarhus University, Gustav Wieds Vej 14, Aarhus, Denmark

[4]Department of Physics, Tamkang University, Tamsui 251, Taiwan

[5] Symbiosis Center for Nanoscience and Nanotechnology (SCNN), Symbiosis International (Deemed University) (SIU), Lavale, Pune, 412115, Maharashtra India

[6]National Center for Nanotechnology and Nanoscience, University of Mumbai, Kalina Campus Kalina, Santacruz (E), Mumbai 400 098

[7] Deutsches Elektronen-Synchrotron, Photon Science, Notkestr, 85, 22607 Hamburg, Germany,

[8]Physics Department, Oakland University, Rochester, Michigan 48309-4401

**Abstract:** We have observed enhancement and shift in the spin reorientation transition temperature as a consequence of coexistence of orthorhombic and hexagonal phases and higher aspect ratio in nanoscale LuFeO$_3$. Nanoparticles and nanofibers of LuFeO$_3$ are considered for this work. Nanoparticles have 75 % orthorhombic phase and 25 % hexagonal phase, while nanofibers have 23% orthorhombic phase and 77%-hexagonal phase. Larger aspect ratio in case of nanofibers is seen to help strain-stabilize the hexagonal phase in the material. Magnetic measurements show significant difference in the magnetic behavior and spin reorientation temperature; 183K for the nanoparticle case and 150K for the case of nanofibers. Moreover, the ferromagnetic moment is two order of magnitude higher for nanofibers than that of nanoparticles, In hexagonal phase, frustration of triangular lattice, works against the long range ordering while magnetic anisotropy works in favor of the long range ordering, which contributes towards the enhanced and anomalous magnetic behavior in case of fibers. X -ray absorption near edge spectroscopy (XANES) at the Fe K-edge has been used to probe the symmetry driven dynamics of Fe 3d- 4p orbitals. It established that due to noncentrocymmetry of the Fe atom, the nanofibers have decreased 3d-4p orbital mixing and reduced crystal field splitting energy, which are also contributing factor for the enhanced magnetic behaviour.

The physical properties and functional performance of a nanoscale multifunctional material depend upon the morphology, dimensionality, interphase, symmetry, and local structure in the lattice. These can thus also be used as tools to tune the physical properties of such materials. Nanostructures with large aspect ratio are one of the key traits to achieve novel (and in some cases enhanced) functionality and are relevant to next generation scaled down device architectures. For this very reason, one-dimensional nanostructures such as nanowires, nanotubes, nanorods, nanofibers etc., have been widely investigated in recent years.[1–3] Another key factor which potentially brings out uncommon application-worthy behaviours is the interface, since it has a crucial role in controlling and unveiling the interesting functionalities of materials. Indeed, a wide range of physical phenomena such as magnetism, ferroelectricity, multiferroicity, superconductivity and magnetoelectricity have their origin and control at interfaces via epitaxy, strain, reconstruction, polarity, dynamics of spins and orbitals, and electronic band alignments.[4] Interfaces between two distinct structural phases of the same material are even more exciting. Herein we address such an interesting situation pertaining to Lutetium orthoferrite ($LuFeO_3$) in the context of its magnetism, which is of considerable interest at this time.

Lutetium orthoferrite ($LuFeO_3$) is an oxide material with emergent interest in the intensely investigated field of multiferroicity. This is an extraordinary material, which exhibits both orthorhombic (o-) and hexagonal (h-) structures.[5–7] There is a significant difference in lattice symmetry and coordination of Fe ion in the case of these o- and h-structures. In o-LFO, ferroelectricity is unexpected due to the symmetry of lattice; while a canting of Fe moments towards c-direction gives rise to weak ferromagnetism below the Neel Temperature (620 K). On the other hand, weak ferromagnetism is forbidden in h-LFO unless the moments are along the a-axis.[8] The crystal structure of o-LFO is distorted perovskite-type assigned to Pbnm space group. In this structure $Fe^{3+}$-$Fe^{3+}$ super-exchange interaction is very strong resulting in high Néel temperature $T_N \sim 620$ K. In o-LFO, there are no $Lu^{3+}$-$Lu^{3+}$ or $Lu^{3+}$-$Fe^{3+}$ interactions as $Lu^{3+}$ has no localized magnetic moment.[9] Weak ferromagnetism (WFM) exists in o-LFO due to Dzyaloshinskii-Moriya (D-M) interaction.[10] The WFM originates from the anisotropy and antisymmetric spin coupling (ASC).[9] o-LFO has the largest structural anisotropy and the strongest ASC and WFM in the $RFeO_3$ (R=Rare Earth) family. The crystal structure of h-LFO belongs to polar P63cm space group; the Néel temperature for which is $T_N \sim 155$ K and Curie temperature $T_C \sim 1020$ K.[11]

LuFeO$_3$ has been explored by many groups in its hexagonal ceramic and thin film forms[11–17] and in the orthorhombic form.[18–21] Recently, Song et al. reported hexagonal-orthorhombic morphotropic phase coexistence and hexagonal to orthorhombic phase transition in LuFeO$_3$ thin films.[5,6] Cao *et. al.* have reported the electronic structure for the conduction bands of both hexagonal and orthorhombic LFO thin films using x-ray absorption spectroscopy at oxygen K edge. It is reported that, while the oxidation states do not change, the spectra are sensitive to the changes in the local environments of the Fe$^{3+}$ and Lu$^{3+}$ sites in the h- and o- structures. We have also recently reported strain-stabilization of different proportions of -o and -h phases in LuFeO$_3$ nanoparticle and nanofiber systems[7] and its consequences for extraordinary polarization and nanogenerator properties. In this work we examine the consequences of the co-existence of o-LFO and h-LFO phases in different proportions in two different nanoscale morphologies i.e. nanoparticles (NPs) and nanofibers (NFs) on the magnetic behaviour. Indeed, the high concentration of interfaces in nanoparticles provides the grounds for improved coupling of the magnetic phases.[22,23]. As shown herein, we observe significantly different magnetic response and shifted spin reorientation transition in the case of LFO NP and LFO-NF and attribute it to the differing o- and h- contributions in the two cases.

LFO-NPs were synthesized using a sol-gel route combined with post-synthesis annealing.[24] For the synthesis of LFO-NF electrospinning technique was used as discussed previously.[7] The NPs and NFs were annealed up to the temperature just below the temperature of forming the pure orthorhombic phase, to assess the stability of hexagonal phase in different morphologies. In addition to different standard materials characterizations, the investigation of local structure around the transition metal ion (Fe$^{3+}$) was performed by the X-ray Absorption Near Edge Structure (XANES) which reveals the variation of the local structure as the symmetry of the Fe ion changes. This information is key to the magnetic behaviour, since it directly impacts the dynamics of 3d-4p orbitals and crystal field splitting energy. Detailed experimental and characterization specifications are provided in supplementary Information (SI).

Figure 1 (a1-i1) and (a2-i2) show the TEM, HRTEM images (showing the presence of orthorhombic and hexagonal phases), SAED pattern, and elemental mapping in STEM mode for NPs and NFs, respectively. The growth of Pbnm and P63cm phases of LFO-NPs and NFs takes place in such a way that first the P63cm phase nucleates and then the Pbnm phase grows by strain relaxation.

Annealing to higher temperatures makes the material attain the Pbnm phase. We also observed that while going from NP to NF, the Pbnm phase undergoes an increase in strain while the P63cm unit cell slightly relaxes.[7] This shows that the growth model for individual cases of NPs and NFs is quite different from that of the thin film growth model. [5]

Based on the observed angles between the (112) planes of the hexagonal and orthorhombic phases in the TEM images, the two phases are seen to be interfaced at φ such that 90°< φ<120°. The (112) plane is the one with maximal atomic density (and therefore, highest intensity in the Pbnm phase's XRD profile) and near maximal atomic density (not an actual peak in the P63cm phase's XRD profile). This plane could form an interface between the two when rotated by a suitable non-trivial angle. [5]

Intensity normalized Raman spectra for both the samples are shown in Figure 2. The synthesis conditions and ionic radii of rare earth cation governed the phase of rare earth oxides.[25]

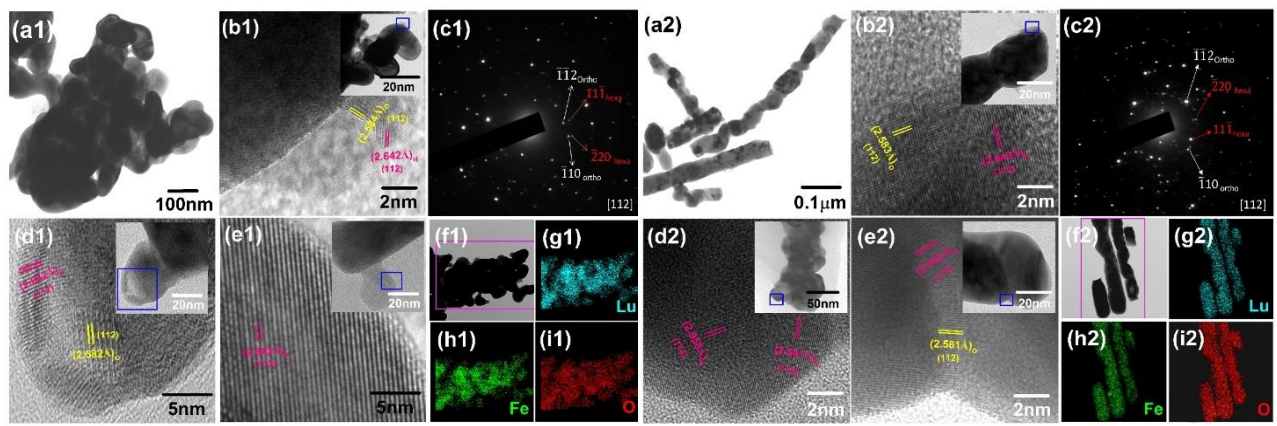

**Figure 1**. TEM image (a1 and a2) HRTEM (b1,b2),(d1,d2) and (e1,e2) showing presence of orthorhombic and hexagonal phases and (c1,c2)SAED and mapping in STEM mode(f1,f2)-(i1,i2) respectively for NP and NF.

According to group theory o-LFO phase exhibits $7A_g+5B_g+7B_{2g}+5B_{3g}$ Raman active modes, while h-LFO phase shows $9A_1+14E_1+15E_2$ Raman active modes.[11,16,26] Table 1 shows the observed Raman modes and their attributes for both the cases observed in the scanning range of 100 to 700 cm$^{-1}$. The Raman peak positions for both the samples are in good agreement with the orthorhombic and hexagonal phases of $LuFeO_3$.[11,16,26]

Figure 2 shows that in the case of NP, -o phase is dominant and -h phase appears to be minor, which is exactly opposite in the NF case. The Raman mapping (Figure 2(b) and (c)) of the certain modes is helpful to quantify the phase contribution in the nanostructures. The Raman mapping for both the cases confirms the phase distribution (as established by XRD data refinement), in both the cases and also reveals that both these phases have a phase boundary.[6] Moreover, it indicates that experimental conditions have created favourable conditions for the hexagonal phase of $LuFeO_3$ to grow on the orthorhombic $LuFeO_3$ structure controlling their relative concentration in both the morphologies.[27] Since free energy of formation of o-$LuFeO_3$ is lower than that of h-$LuFeO_3$, this favours the formation of hexagonal phase on the orthorhombic phase to cross the energy barrier over the length scale of sub-micrometer.[6,7]

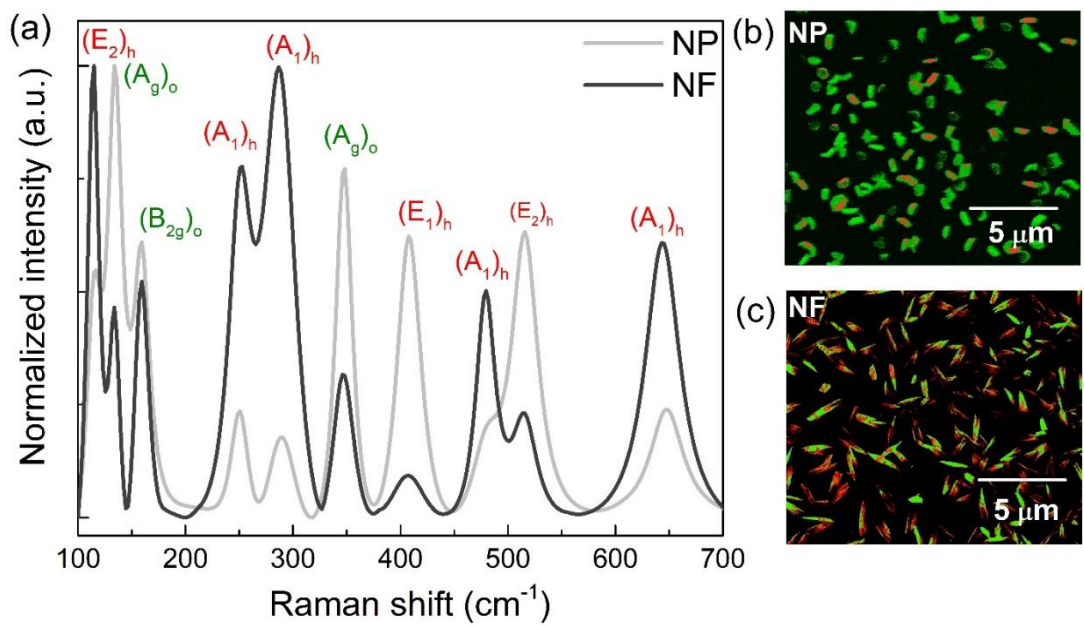

Figure 2. Raman spectroscopy study of LFO NP and NF. (a) normalized Raman spectra for both the cases. The highlighted region of both spectra shows Raman modes considered for Raman mapping, (b) and (d) show Raman mapping image obtained on selected Raman modes for NP and NF cases respectively.

**Table 1**. Observed Raman modes in the present study and their one to one comparison with the reported Raman modes for orthorhombic and hexagonal phases, alongwith their designation.

| Orthorhombic | | | | Hexagonal | | | | Present Study (mixed phase) | |
|---|---|---|---|---|---|---|---|---|---|
| RFO[24] | LFO[26] | mode | Attribute | LFO[1] | YMnO[28] | mode | Direction and sign of largest atomic displacements | NP | NF |
| 110 | 110 | - | **R-Motion** | 110 | - | E2 | +x,y(Fe,O3,O4) -x,y(Lu1,Lu2) | 115 **(h)** | 114 **(h)** |
| 136 | 136 | $A_g(6)$ | **R-Motion** | - | 135 | E2 | +x,y(Lu1) -x,y(Lu2) | 133 **(o)** | 133 **(o)** |
| 160 | 158 | $B_{2g}(5)$ | **R- Motion** | - | 148 | A1 | +z(Lu1) -z(Lu2) | 158 **(o)** | 155 **(o)** |
| - | - | | | - | 190 | A1 | rot. x,y(FeO$_5$) | - | - |
| - | - | | | 214 | 215 | E2 | +x,y(O2,Fe) -x,y(O1,O3) | - | - |
| 225 | | $A_g(2)$ | **FeO$_6$-Rotation** | 223 | 257 | A1 | +z(Lu1,Lu2) -z(Fe) | 252 **(h)** | 251 **(h)** |
| 290 | 278 | $B_{1g}(3)$ | **FeO6-Rotation** | 289 | 297 | A1 | x(Fe), z(O3) | 286 **(h)** | 284 **(h)** |
| 349 | 350 | $A_g(7)$ | **R-O Vibration** | 346 | 302 | E2 | +z(O2) -z(O1), x,y(O4) | 346 **(o)** | 347 **(o)** |
| - | - | | | - | 376 | E1 | +x,y(O1) -x,y(O2) | - | - |
| 409 | 425 | | **R-O vibration** | 404 | 408 | E1 | +x,y(O1) -x,y(O2) | 407 **(h)** | 406 **(h)** |
| - | - | | | - | 433 | A1 | +z(O4,O3) -z(Fe) | - | - |
| - | - | | | - | 459 | A1 | +x,y(O1,O2) -x,y(Fe) | - | - |
| 428 | 427 | $A_g(4)$ | **FeO$_6$ bending** | 448 | 458(c) | E2 | +x,y(O4) -x,y(O1,Fe) | - | - |
| 453 | 450 | $B_{1g}(4)$ | **FeO$_6$ bending** | 473 | - | A1 | | 477 **(h)** | 475 **(h)** |
| 517 | 516 | | | 501 | 515(c) | E2 | +x,y(O4,O3) +x,y(O1,O2) | 516 **(o)** | 514 **(o)** |
| 608 | - | $A_g(3)$ | **FeO$_6$ stretching** | 603 | 632 | E1 | x,y(O3) -x,y(O4) | - | - |
| 644 | 654 | $B_{2g}(2)$ | **FeO$_6$ stretching** | 651 | 681 | A1 | +z(O1) -z(O2) | 644 **(h)** | 646 **(h)** |
| - | - | | $A_g(1)$ **FeO$_6$ stretching** | 721 | - | A1 | | - | - |

For rare earth orthoferrites, the Raman modes below 200 cm$^{-1}$ mainly describe the displacements of rare earth ions, while, motions of iron and oxygen ions are accounted for by the modes above 300 cm$^{-1}$.[11,25,28,29] Thus, changes in the Lu - O and Fe - O bond lengths and FeO$_5$ bipyramidal tilt ultimately lead to shifts in the Raman modes.

Figures 3 (a) and (b) show the results of the temperature dependent magnetic moment measurements from room temperatures down to 2 K for NPs and NFs, respectively. The development of a ferromagnetic component is observed from $T_{SR2}$ =183 K to $T_{SR1}$ = 153K in the case of NPs and from $T_{SR2}$ =150 K to $T_{SR1}$ = 130K in the case of the NFs. This indicates evolution of a second magnetic phase in the material by spin reorientation transition.[12]

The Neel Temperature of o-LFO is reported to be $T_{N(o)}$ =620 K,[12] while for h-LFO different values have been reported. Wang et al. reported $T_{N(h)}$ = 440K in 2014[30], which is much higher than the value of 155 K subsequently reported by Disseler.[11]

Figure 3 (e) and (f) show the crystal lattice of orthorhombic and hexagonal LFO. The triangular spin lattice and magnetic anisotropy impact the ordering temperature in h-LFO significantly. The triangular spin lattice in h-LFO creates the frustration and contributes towards lowering down the long range spin ordering temperature.[31–34] Simultaneously, the magnetic anisotropy energy in the crystal works against the frustration of the triangular lattice and adds to long range magnetic order, which also affects the spin reorientation in the system.[33,34] This spin frustration due to triangular lattice in h-LFO introduces interesting magnetic phases. Below the Neel temperature $T_{N(h)}$ = 440 K, magnetic order in h-LFO emerges again by a spin reorientation resulting in weak ferromagnetism due to the D-M interaction and single-ion anisotropy mechanism.[30]

In case of o-LFO, the frustration in the lattice is lower than that of h-LFO, hence magnetic anisotropy does exist. The presence of magnetic anisotropy adds to the long-range spin order, which in turn favors a higher value of magnetic ordering temperature as well as spin reorientation temperature. This explains the higher magnetic transition temperatures in case of o-LFO than that of h-LFO. The interplay and tuning between the frustration of triangular lattice and magnetic anisotropy play significant role for the high/low magnetic ordering temperature in o/h-LFO.

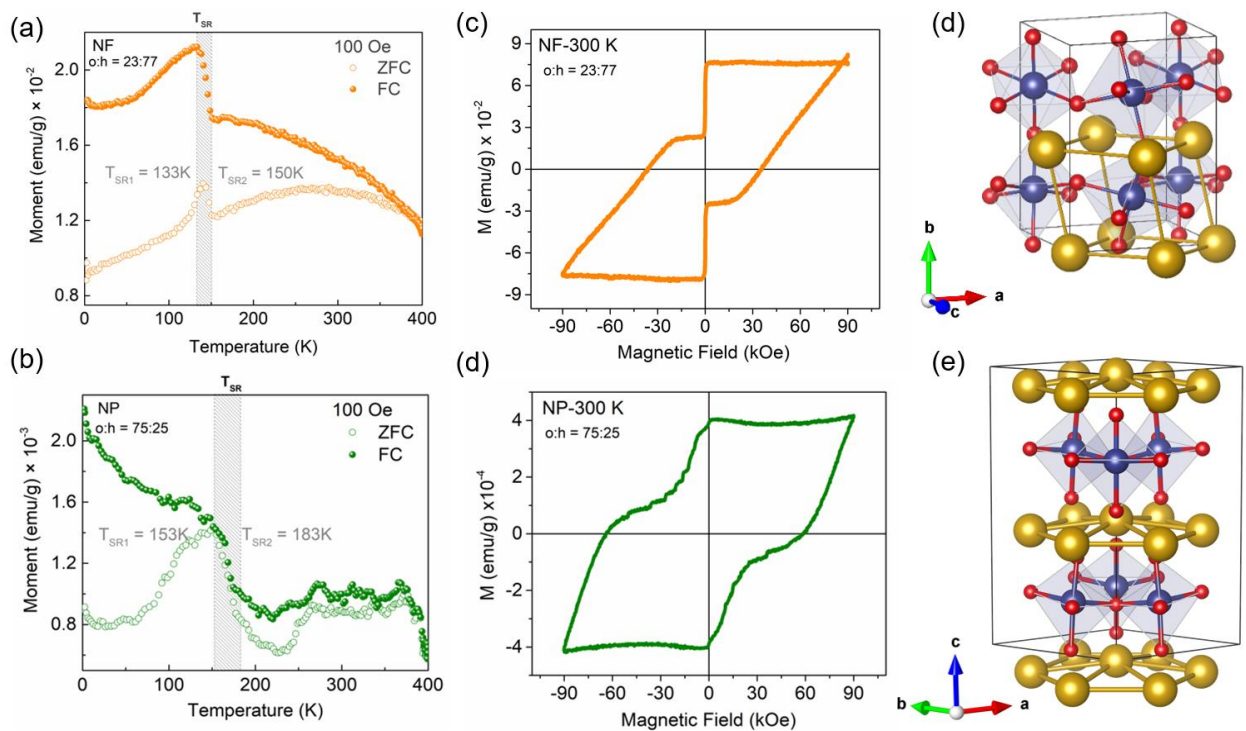

**Figure 3**. (a-b)Temperature dependent magnetization for temperatute range 300 K to down 0 K for NF and NP respectively, (c-d) Ferromagnetic component (after paramagnetic/diamagnetic background removed) of MH hysterisys loop for NF and NP for 300 K and (e-f) orthorhombic lattice and hexagonal latticescheamatic.

To find out the intrinsic magnetic behaviour of mixed phase o-h LFO, we measured the M-H hysteresis. Figure 3(c-d) show ferromagnetic M-H loops at 300 K, obtained by removing the paramagnetic and diamagnetic backgrounds. (M-H loops for 300k and 5 K without background subtraction are shown in SI Fig S2) Distinctly LFO NF show two orders higher magnetization value than that of NP. Also the double-step behaviour of the loop is more clearly defined in the case of LFO NF (which has dominantly the h-phase) is similar to the magnetization data reported by Moyer et al. and Wang et al. for thin films of h-LuFeO$_3$ phase.[27,35] They have suggested that the first step at H=0 is possibly coming from small amount of iron rich impurity phase. In our case since the X ray diffraction pattern and refinement data do not show any possibility of impurity phase, the possibility of interplay and coupling between the interfaces of o- and h- phases (with soft phase depinning at low field) and the dynamics of spins/lattice degrees of freedom are possibly responsible for such behaviour. Cao et al. have reported phase separation in LuFeO$_3$ thin films, wherein the boundary between the o- and h- structural phases show a tendency to align with the crystal planes of the h-LuFeO$_3$ phase.[6] The interface exchange and coupling may also be the cause of the enhancement of magnetic moment in case of fibers as compared to nanoparticles. This, compounded with strong shape anisotropy would render the differing character to the two hysteresis loops shown in Fig. 3 (c-d).

The refinement of XRD (figure S1 in SI) data shown in Table 1 provides the interatomic distances of Fe-O, for FeO$_6$ and FeO$_5$ polyhedra.[7] NF have smaller Fe-O-Fe bond lengths than NP. The magnetic ordering of the Fe-O-Fe superexchange is sensitive to the bond angle and bond length.[36] The Fe-O-Fe superexchange angle is reduced in the case of NFs, which is also one of the favoring factor for enhanced magnetic moment. These observations and analysis are in coherence with the recent report by Suresh et al. wherein they have reported the coupling between lattice, electric and magnetic degrees of freedom in the case of bulk h-LFO based on neutron diffraction and x ray diffraction.[37] They have also suggested the presence of strong magneto-elastic coupling, which would be stronger in the NF case.

**Table 1.** Critical structural parameters obtained Rietveld refinement of by X ray diffraction data

| | Pnma (o-Phase- FeO6) | | | | | | | | |
|---|---|---|---|---|---|---|---|---|---|
| | a | b | c | V | % | Fe-O | FeO6 | <Fe-O-Fe> | φ |
| NP | 5.55 | 7.56 | 5.21 | 218.54 | 75 | 2.99 | 10.71 | 144.847 | 17.58 |
| NF | 5.56 | 7.57 | 5.22 | 219.36 | 23 | 3.04 | 10.79 | 144.327 | 17.84 |
| | P63cm (h- phase- FeO5) | | | | | | | | |
| | a | b | c | V | % | Fe-O | FeO5 | <Fe-O-Fe> | φ |
| NP | 5.93 | 5.93 | 11.77 | 358.53 | 25 | 3.26 | 7.26 | 118.932 | 30.53 |
| NF | 5.95 | 5.95 | 11.73 | 359.50 | 77 | 3.3 | 6.55 | 117.303 | 31.30 |

Further, as reported in our recent work,[25] The biphasic NFs with higher h/o ratio (77:23) exhibit higher electric polarization as compared to the NPs with lower h/o ratio (75:25) at room temperature. By virtue of mutual exclusion of polarization and ferromagnetism/antiferomagnetism cannot simultaneously exist within a single-phase at RT, the observed magnetism and polarization in the o-h LFO possibly can be caused by the magnetoelectric coupling/interaction between the o -LFO and h -LFO phases at the boundaries similar to that reported by Song et al. in case of Morphotropic o-h phase coexistence in LuFeO3 thin films.[5] The coexistence of o and h phase is responsible also for the observed shift in the magnetic reorientation transition in case of NF through magnetoelectric coupling effects.

To investigate further, the difference in magnetic behaviour and interaction between Lu and Fe lattices in NP and NF, we investigated the local structure of the samples. The interplay between spin lattice frustration and magnetic anisotropy energy is a consequence of the local structure in the system around the Fe atom. Spin frustration in such system can present extraordinary magnetic phases.[38] Synchrotron XANES at the Fe K edge corresponds to the electronic transitions from 1s to 4p states and is sensitive to the octahedral environment of Fe atom and local electrostatic interaction.[39] The symmetry of the transition metal site affects the XANES spectra of transition metal oxides significantly. Symmetry around the absorbing atom affects the pre-edge transition. The displacement from centrosymmetry induces the mixing of p-orbital from the neighbouring oxygen octahedron with the d-orbital of the iron atom. The appearance of a large peak in the XANES spectrum of transition metal, before the main peak (pre-edge) in the spectrum indicates the possibility of significant displacement of the iron atom from centrosymmetry.[40,41] The decreasing coordination number in Fe site indicates lack of inversion symmetry at the iron site. This gives rise to the 1s-3d (pre-edge) transition peak. The low symmetry phase showing stronger pre-edge intensity indicates the hybridization of metal 4p and 3d orbitals, which in turn provides allowed electric-dipole 1s-4p transition.[41] Due to much weaker quadrupole coupling than the electric-dipole coupling, the hybridization of metal 4p and 3d orbitals may significantly affect the 1s-3d pre-edge transition.[41] The Fe K-edge absorption (main peak) is primarily due to the the electric-dipole coupling 1s-4p transition.[39,41] Splitting of the pre-edge spectral peaks is a indication of the d-orbital splitting of the half-filled $Fe^{3+}$ d-states into two energy levels comprising of three degenerate nonbonding $t_{2g}$ and the two degenerate anti-bonding $e_g$ orbitals. The energy difference between the $t_{2g}$ and $e_g$ orbitals gives the octahedral crystal field splitting energy ($\Delta_o$). It is a significant spectral factor which determines the spin configuration of the transition metal complex, and therefore its magnetism.[39] The difference in the maxima of the A doublet peak gives the value of $\Delta_o$. The reduced field splitting energy indicates higher spin state, reduced hybridization of the O 2p and Fe 3d states and hence enhanced magnetization.[41]

In the case of LFO NP and NF we recorded the temperature dependent Fe K edge XANES spectra. Figure 4(a) shows the pre edge and main-edge XANES spectra for NP and NF for temperature range 20 K to 300 K, which were measured at the XAFS-beamline P64 at PETRA III, DESY[42]. Figure 4 (b) shows zoomed out view of pre-edge A and its splitting for NPs and NFs for various temperatures. Figure 4(c) shows the $FeO_6$ ($FeO_5$) polyhedra belonging to orthorhombic (hexagonal) symmetry. Since NFs dominantly have hexagonal phase, it possesses non-centrosymmetry and lower coordination number of Fe. LFO NPs and NFs experience different amounts of distortion, owing to the difference in symmetry, morphology, and coordination of Fe atoms

. This is observed experimentally in terms of the difference in the signature of their pre-edge XANES spectra. The decreased coordination number due to non-centrosymmetry ($FeO_5$) in NFs is confirmed by the higher intensity of pre-edge of NFs for all the temperatures as compared to that of NP, which has dominant centrosymmetric attribute due to the presence of $FeO_6$ polyhedra in majority.

The crystal field splitting energy $\Delta_o$ of the d orbital of Fe atom is calculated using the absorption profile of pre-edge feature of the samples. The crystal field splitting energy values with their respective values of temperature for NPs and NFs are shown in Table 2. Lower value of $\Delta_o$ in the case of NFs as compared to that of NPs at all the temperatures beyond spin reorientation can be attributed to the increase in the number of occupied states in the Fe 3d orbital in NF. Hence due to higher spin state, the hybridization of the O 2p and Fe 3d states is reduced and higher magnetization is observed.[41] Further, individual trends of $\Delta_o$ also corroborate with the aforesaid relation between $\Delta_o$ and the magnetic behaviour.

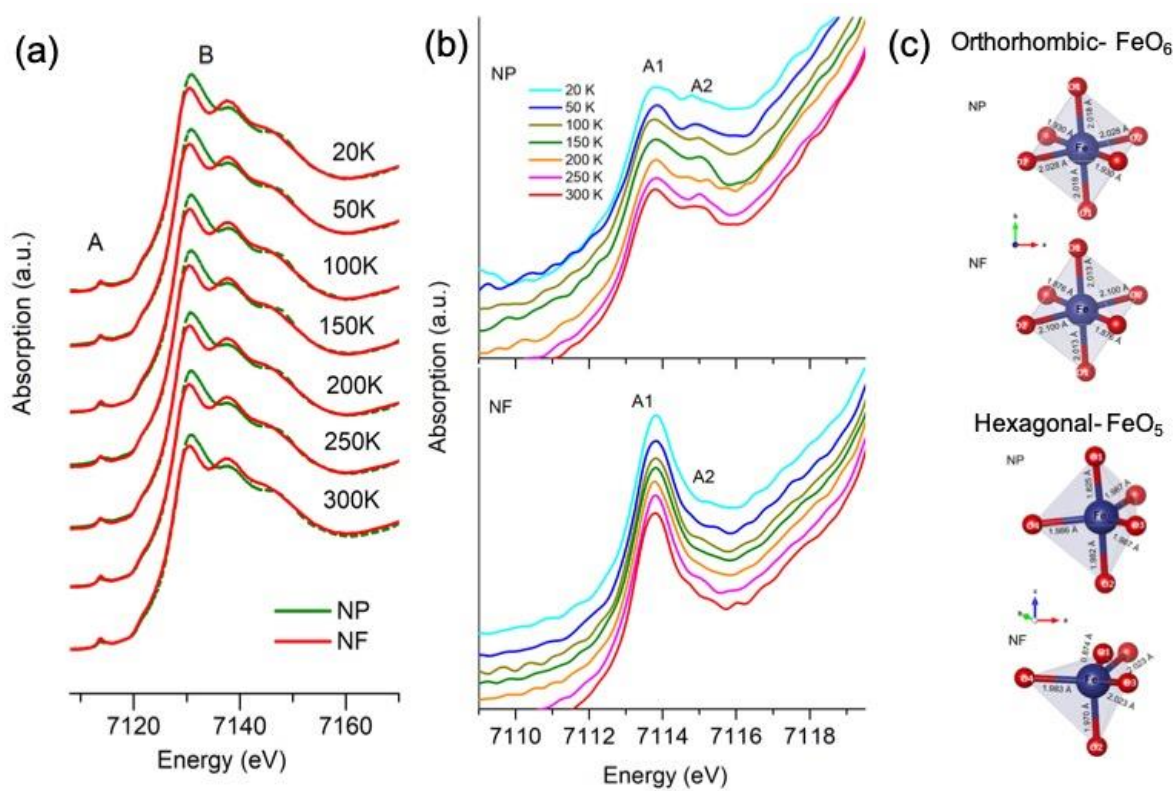

**Figure 4.** (a) Temperature dependent XANES Pre-edge (Peak A) and main peak for LFO-NP and LFO-NF, (b) splitting of pre-edge for temperatures 20,50,100,150,200, 250 and 300 K and (c) FeO6 and FeO5 polyhedra in case of orthorhombic and hexagonal phase respectively based on bond lengths calculated from Reitveld refinement of the XRD data.

Table 2. Calculated crystal field splitting Energy.

|     | NP | | | NF | | |
| --- | --- | --- | --- | --- | --- | --- |
|     | t2g | eg | $\Delta_o$ | t2g | eg | $\Delta_o$ |
| 20  | 7114.91 | 7113.58 | 1.33 | 7115.43 | 7113.75 | 1.68 |
| 50  | 7114.99 | 7113.72 | 1.27 | 7115.45 | 7113.74 | 1.71 |
| 100 | 7115.45 | 7113.73 | 1.72 | 7115.21 | 7113.75 | 1.46 |
| 150 | 7115.13 | 7113.55 | 1.58 | 7115.34 | 7113.76 | 1.58 |
| 200 | 7115.21 | 7113.68 | 1.53 | 7115.3 | 7113.72 | 1.58 |
| 250 | 7115.08 | 7113.61 | 1.47 | 7115.1 | 7113.75 | 1.35 |
| 300 | 7115.07 | 7113.64 | 1.43 | 7114.8 | 7113.72 | 1.08 |

In summary, LuFeO$_3$ NPs and NFs demonstrate distinctly different magnetic behaviours. Since NF dominantly has hexagonal phase (LFO-NF: 23%-o and 77%-h), the corresponding ordering temperature is lower than that of the NPs (LFO-NP: 25%-o and 75%-h), since it is driven differently by the dynamics between triangular spin lattice frustration in hexagonal lattice and magnetic anisotropy energy in orthorhombic lattice. Possibility of the influence of the differences in the spin orbit coupling between the NF and NP cases due to the significant differences between the o- and h-contents therein cannot be ruled out. Enhanced magnetization in NFs is governed by local structure of Fe ion. Non-centrosymmetry and lower coordination number facilitate decreased crystal field splitting energy for NFs, which promote higher spin state therein and reduced hybridization of the O2p and Fe 3d states, which results in enhanced magnetization.

**Acknowledgements**:

This work was carried out under the Grant No. SR/WOS-A/PM-16/2016 from the Department of Science and Technology, Ministry of Science and Technology, India, and a Fulbright Fellowship Grant No. 2372/F-N APE FLEX/2018 availed by S.C. The research at Oakland University was supported by grants (DMR-1808892; ECCS-1923732). The authors also acknowledge the TEM measurements facility provided by SAIF IIT Bombay India. We acknowledge DESY (Hamburg, Germany), a member of the Helmholtz association HGF, for the provision of experimental facilities. Parts of this research were carried out at PETRA III. SBO acknowledges funding support from DST Nanomission under the Thematic Unit program and the Department of Atomic Energy for the Raja Ramanna Fellowship award.

**Table of Content Graphic:**

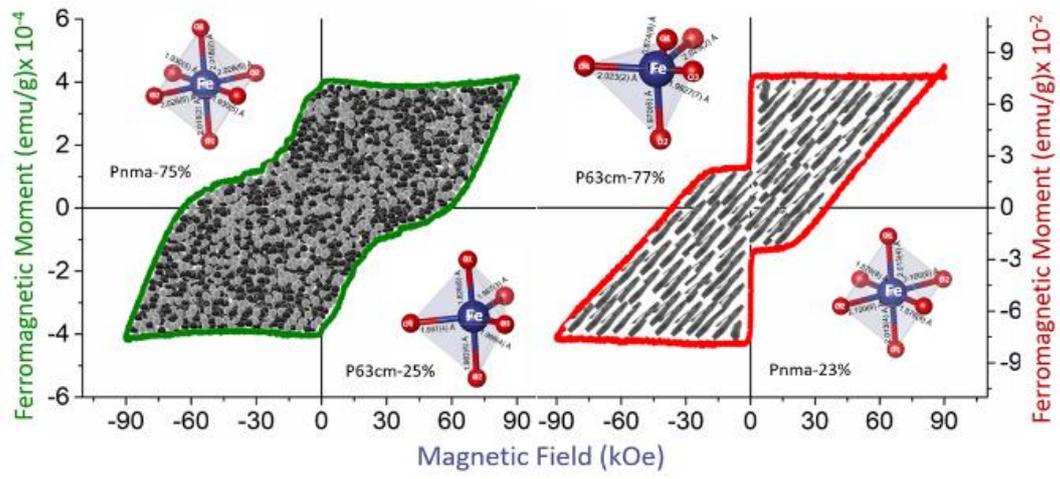